\documentclass{iopart}
\usepackage[varg]{txfonts}   
\usepackage{graphics}
\usepackage[utf8x]{inputenc}
\usepackage{iopams,graphicx,amssymb,textcomp}

\eqnobysec

\begin{document}

\title[Effects of turbulent environment on the surface roughening]
{Effects of turbulent environment on the surface roughening: The
Kardar-Parisi-Zhang model coupled to the stochastic Navier-Stokes equation}

\author{N~V~Antonov$^1$, N~M~Gulitskiy$^1$, P~I~Kakin$^1$ and M~M~Kostenko$^{1, 2}$}
\address{$^1$ Department of Physics, Saint Petersburg State University, 7/9 Universitetskaya nab., Saint~Petersburg 199034, Russia\\
$^2$  L.D.~Landau Institute for Theoretical Physics, 142432, Ak.~Semenova 1-A, Chernogolovka, Moscow region, Russia}
\eads{\mailto{n.antonov@spbu.ru}, \mailto{n.gulitskiy@spbu.ru}, \mailto{p.kakin@spbu.ru}, \mailto{m.m.kostenko@mail.ru}}

\begin{abstract}
The Kardar-Parisi-Zhang model of non-equilibrium critical behaviour (kinetic surface roughening) with turbulent motion of the environment taken into account is studied by the field theoretic renormalization group approach. The turbulent motion is described by the stochastic  Navier-Stokes  equation  with  the random stirring force whose correlation function includes two terms that allow one to account both for a turbulent fluid and for a fluid in thermal equilibrium. The renormalization group analysis performed in the leading order of perturbation theory (one-loop approximation) reveals six possible types of scaling behaviour (universality classes). The most interesting values of the spatial dimension $d=2$ and~$3$ correspond to the universality class of a pure turbulent advection where the  nonlinearity of the Kardar--Parisi--Zhang model is irrelevant.
\end{abstract}
\vspace{2pc}
\noindent{\it Keywords\/}: surface roughening, non-equilibrium critical behaviour, turbulent advection, renormalization group.

\section{Introduction}

The problem of random growth phenomena and fluctuating surfaces has been attracting constant attention over the past few decades~\cite{KPZ}~--~\cite{Xia}.
One of the widely accepted models for those phenomena is provided by the Kardar-Parisi-Zhang (KPZ) stochastic differential equation. It was introduced in~\cite{KPZ}
 to explain universal scaling behaviour observed in various systems with random surface growth.\footnote{To be precise, an equivalent model was introduced earlier in~\cite{FNS} as a stochastic Burgers equation for the purely potential vector field 
 ${\bf v} = {\bf \partial} h$.}
Such scaling behaviour is sometimes referred to as kinetic roughening~\cite{rost1} because the growing surfaces become increasingly rough over time. Examples include dynamics of flame fronts, cancer tumours, bacterial colonies, spreading of cholera and other epidemics, earthquakes, social disturbances, etc. As a result, the KPZ equation became a paradigmatic model of general non-equilibrium critical phenomena.

The scaling of kinetic roughening is described by the power law for correlation functions asymptotic behaviour in the infrared (IR) range (large temporal $t$ and spatial $r$ differences when compared with characteristic microscopic scales)~\cite{rost1}~--~\cite{rost3}:   
\begin{equation}
\langle\left[h(t,{\bf x}) - h(0,{\bf 0})\right]^{n}\rangle \simeq
r^{n\chi}\, F_n (r/t^{1/z}), \quad  r=|{\bf x}|.
\label{scaling}
\end{equation}
The roughness exponent $\chi$ and the dynamical exponent $z$ define the universality class of the scaling behaviour. Here $F_n(\cdot)$ are certain universal scaling functions, the averaging~$\langle\dots\rangle$ is performed over the statistical ensemble, and $h(t,{\bf x})$ stands for the height of the surface profile (here and below, $t$ and ${\bf x}$ are the time and the space coordinates).

However, despite relative simplicity of the formulation of the KPZ model, its paradigmatic reputation,  and numerous attempts made, a satisfactory theoretical understanding of the model has not yet been achieved.
The value of the upper critical dimension for the KPZ model, and its very existence, is still disputed~\cite{LK}~--~\cite{UpperInf}. Even the question of whether the model is well-posed in a rigorous mathematical sense still remains a matter of discussion~\cite{Hairer,Kupiainen}. Moreover, numerous mappings onto various reaction-diffusion models raise questions about the meaning of imaginary random noise~\cite{Howard}~--~\cite{Cooper}. The numerical simulations unavoidably deal with the discretized models instead of the original continuous model. 
Thus, it is not clear whether they belong to the same universality class of IR behaviour; see, e.g.,~\cite{Wiese8} and references therein.

The perturbative renormalization group (RG) analysis within the $\varepsilon$ expansion (where $\varepsilon$ is a deviation from the logarithmic spatial dimension) proved to be extremely successful in the study of numerous equilibrium and non-equilibrium critical phenomena. In the case of the KPZ model, it shows that there is no IR attractive fixed point  in the physical range of the parameters~\cite{11,111}. 
The existence of the strong-coupling nonperturbative fixed point 
was established within the functional RG 
\cite{Canet}~--~\cite{Canet4}, but it has not been confirmed by any other approach based on a systematic expansion in some (at least, formal) parameter. 

All of these facts suggest that the KPZ model might not be self-contained and, instead, might be very sensitive to various extensions, disturbances and modifications. For example, the simple modification proposed by~\cite{Pav} immediately led to a model with infinitely many coupling constants~\cite{Pav2}. Another example was encountered in cosmological applications of the KPZ model, where it was applied to description of self-gravitating uniform medium~\cite{Uni}~--~\cite{UniA}. Inclusion of  non-potential degrees of freedom produced a typical IR attractive fixed point within the corresponding $\varepsilon$ expansion~\cite{UniA}. 

Experience with nearly-equilibrium nearly-critical systems suggests that they can be drastically affected by the motion of the constituting or surrounding medium. Indeed, the critical scaling behaviour can be destroyed in favour of the mean-field behaviour or completely new non-equilibrium universality classes~\cite{Onuki2}~--~\cite{Nelson}.
Thus, it is highly desirable to study the effects of the medium motion (which can hardly  be excluded in real experimental settings) on the critical behaviour of fluctuating surfaces.  Two such attempts were undertaken recently. In~\cite{AKL}, the KPZ model was coupled to the stochastic Navier-Stokes (NS) equation driven by a short-correlated random noise (which corresponds to the fluid in thermal equilibrium), an ensemble proposed in~\cite{FNS}. No physically acceptable nontrivial fixed points were found within the perturbative RG analysis.

In~\cite{Us}, the velocity was modelled by a ``synthetic'' turbulent Gaussian ensemble with vanishing correlation time, known as the Kazantsev-Kraichnan ensemble; see~\cite{FGV} and references therein. It was shown that for incompressible case and for the most interesting values of the spatial dimension $d=2$ and~$3$, the KPZ nonlinearity appears IR irrelevant in the sense of Wilson, that is, the IR scaling behaviour is completely determined by the turbulent advection.

Of course, it is desirable to consider more realistic velocity ensembles, in particular, to include finite correlation time. Unfortunately, synthetic Gaussian ensembles with finite correlation time suffer from the lack of Galilean symmetry, which plays an important role in the whole problem.

In this paper, we employ the stochastic NS equation for the incompressible viscid fluid,
which implies finite correlation time, non-Gaussianity and, at the same time, is manifestly Galilean covariant. Moreover, now the velocity field has its own dynamics, which opens the possibility to study the feedback of the advected fields on the fluid dynamics itself.

The random stirring force has a power-like correlation function, namely, 
\mbox{$\propto \delta(t) \,  k^{4-d-y}$}, where $k$ is the wave number, $d$ is the spatial dimension and $y$ is an exponent with the logarithmic value $y=0$ and the physical value $y\to4$.
This  choice is  typical for the standard RG treatment of the problem~\cite{Red,Vasiliev}. 
For renormalizability reasons, the correlation function should be modified by inclusion of the local in-space term \cite{HoNa,AHKV}. This extended model allows one to consider  both the turbulence and the fluid in thermal equilibrium. 

We apply to the problem the field theoretic RG. 
In this approach, possible types of IR asymptotic behaviour are associated with IR attractive fixed points of the corresponding RG equations.
Practical calculations are performed in the leading one-loop approximation, but the critical exponents 
in the scaling relation (\ref{scaling}) are  found exactly.

Our main result is as follows: for the turbulent environment ($y\to4$) and the most interesting physical values $d\ge2$ the nonlinearity of the KPZ equation becomes irrelevant and the IR asymptotic behaviour is described by pure turbulent advection.  Although rather disappointing, this result seems reliable and agrees with the one derived earlier in \cite{Us} for a simpler velocity ensemble.

The plan of the paper is as follows: detailed description in the model and its field theoretic formulation age given in sections~2 and~3, respectively. The RG analysis, one-loop RG functions and RG fixed points are discussed in section~4. Section~5 is reserved for the conclusion.

\section{The KPZ model and the NS equation \label{smod}}

The KPZ equation is a semi-phenomenological model described by the  stochastic differential
diffusion-type equation 
with the simplest nonlinear term that respects the symmetries $h\to h+$const and $O(d)$:
\begin{equation}
\partial_{t} h=\kappa_{0} \boldsymbol{\partial}^{2} h+\lambda_{0}(\boldsymbol{\partial} h)^{2}/2+f. 
\label{KPZ1}
\end{equation}
Here $\kappa_{0}$ and $\lambda_{0}$ are the coefficients of surface tension and lateral growth respectively, while $f=f(t,{\bf x})$ is a random noise that represents small-scale perturbations.
Here and below we denote $\partial_{t}= \partial/\partial t$,
$\boldsymbol{\partial}= \{\partial/\partial x_{i}\}$, 
$\boldsymbol{\partial}^{2}=(\boldsymbol{\partial}\cdot\boldsymbol{\partial})$, 
$(\boldsymbol{\partial} h)^{2}=(\boldsymbol{\partial} h\cdot\boldsymbol{\partial}h)$.

The statistics of $f$ is implied to be Gaussian with a zero mean and the pair correlation function
\begin{equation}
\langle f(t, {\bf x})f(t',{\bf x}') \rangle = C \delta(t-t')\, \delta^{(d)}({\bf x}-{\bf x'}),
\label{covar}
\end{equation}
where the positive amplitude can be set to $C=1$ without loss of generality.

To include the advection by turbulent environment, the equation~(\ref{KPZ1}) should be modified by the ``minimal'' replacement
of the ordinary time derivative $\partial_t$ with its Galilean covariant counterpart (Lagrangean derivative) $\partial_t \to \nabla_t=\partial_t + (\boldsymbol{v} \cdot\boldsymbol{\partial})$: 
\begin{equation}
\nabla_t h  =\kappa_{0} \boldsymbol{\partial}^{2} h+\lambda_{0}(\boldsymbol{\partial} h)^{2}/2+f.
\label{KPZ}
\end{equation}
The the velocity field $\boldsymbol{v}$
is described by the stochastic NS equation for an incompressible viscous fluid: 
\begin{equation}
\nabla_{t} \boldsymbol{v}
 = \nu_0 \boldsymbol{\partial}^2 \boldsymbol{v} - \boldsymbol{\partial} \wp + \boldsymbol{F}.
\label{NS}
\end{equation}
Here $\wp$ is the pressure, $\boldsymbol{F}$ is the transverse external random force, $\nu_0$ is the kinematic viscosity coefficient; both $\boldsymbol{v}$, $\wp$, and $\boldsymbol{F}$ depend on $\{t, {\bf x}\}$.
The field $\boldsymbol{v}$ is transverse due to the incompressibility condition $(\boldsymbol{\partial}\cdot \boldsymbol{v})=0$. The external stirring force $\boldsymbol{F}$ has a Gaussian statistics with a zero mean and the given correlation function:
\begin{eqnarray}
\langle F_{i} (t, {\bf x}) F_{j}(t',{\bf x}')\rangle =  \delta(t-t')\,\int\,
\frac{d{\bf k}}{(2\pi)^{d}} \,
\ P_{ij}({\bf k})\, D(k)\,
e^{{\rm i} {\bf k}\cdot ({\bf x-x'})},
\label{white}
\end{eqnarray}
where $P_{ij}({\bf k}) = \delta_{ij} - k_i k_j / k^2$ is the transverse 
projector, $k\equiv |{\bf k}|$ is the wave number. \footnote{The equations~(\ref{KPZ}), (\ref{NS}) are studied on the entire $t$ axis and are supplemented by the retardation condition and by the condition that the fields vanish asymptotically for $t\rightarrow -\infty$. The $\delta$-function in-time in (\protect\ref{white}) ensures the Galilean symmetry.}

The function $D(k)$ in~(\ref{white}) is usually chosen in the power-like form
\begin{equation}
D(k) = D_{10} \, k^{4-d-y},  \quad D_{10}>0,
\label{DNS}
\end{equation}
typical for the standard field theoretic approach to the fully developed turbulence; see, e.g., the monographs~\cite{Red,Vasiliev} and references therein. 
The physical value of the exponent $y$ corresponds to the limit $y\to4$, when the function~(\ref{DNS}) with the proper choice of the amplitude $D_{10}$
can be viewed as a power-like representation of the function $\delta({\bf k})$ that describes the energy pumping by large-scale stirring.

The model~(\ref{NS})~--~(\ref{DNS}) is logarithmic (the corresponding coupling constant $g_{10}=D_{10}\nu_0^{-3}$ is dimensionless)
at $y=0$ and arbitrary $d$; the ultraviolet (UV) divergences in the perturbation theory have the forms of the poles in $y$. 
However, the KPZ model~(\ref{KPZ1})~-- (\ref{covar}) becomes logarithmic at $d=2$, 
and its RG analysis should be performed within the expansion in $\varepsilon=2-d$. In order to make the RG analysis of the full model internally consistent, it is necessary to treat $y$ and $\varepsilon$ as small parameters of the same order.\footnote{Otherwise one of the interactions should be neglected from the very beginning as IR irrelevant in the sense of Wilson and some nontrivial asymptotic regimes would be lost.}
Then the UV divergences take on the form of the poles in $y$, $\varepsilon$ and their combinations, while the coordinates of the fixed points and various critical dimensions are calculated as double series in  $y$ and $\varepsilon$.

In its turn, the RG analysis of the model (\ref{NS})~--~(\ref{DNS})
near $d=2$ becomes rather delicate. It shows that, in order 
to ensure the multiplicative renormalizability, it is necessary to add to the random force correlation function a local term (an even integer power of the wave number $k$), namely
\begin{equation}
    D(k) = D_{10} \,k^{2+\varepsilon-y}+D_{20}\, k^2,  
    \label{D}
\end{equation}
where both $D_{10}$, $D_{20}$ are positive  
\cite{HoNa}; see also sec.~3.10 in the monograph~\cite{Red}. Detailed discussion of this issue and the two-loop calculations in various renormalization schemes can be found in \cite{AHKV}. 
Similar situation, when a model is logarithmic for arbitrary $d$ but additional UV divergences arise at some exceptional values of $d$, and extension of the model by adding local terms is required, was encountered recently  in \cite{Tomas,Tomas2}.

Therefore, there are four coupling constants in the full model (\ref{covar})~-- (\ref{D}): 
\begin{equation}
g_{10}=D _{10}\nu_0^{-3}, \quad
g_{20}=D _{20}\nu_0^{-3}, \quad
g_{30}= \lambda_{0}\nu_0^{-3/2}, \quad
w_0 = \kappa_0 \nu_0 ^{-1}. 
\label{4CC}
\end{equation}
Although $w_0$ is not an expansion parameter, it is dimensionless and should be treated on equal footing with the  other three couplings.

For the turbulent fluid, which we are interested here, $g_{20} =0$, but its renormalized analog does not vanish, and it is necessary to keep the both terms in (\ref{D}).
It is also worth noting that such an extended model includes the special case $g_{10} =0$ which is closed with respect to renormalization. 
Physically, it corresponds to the model of a fluid in thermal equilibrium, studied earlier in~\cite{AKL}.

\section{Field theory \label{sft}}

According to the general theorem by De Dominisis-Janssen (see, e.g., 
chap.~5 in the monograph~\cite{Vasiliev}) the original stochastic problem~(\ref{covar})~--~(\ref{D})
can be reformulated as a field-theoretic model for the doubled set of fields 
$\Phi=\left\{v_{i}^{\prime},h^{\prime}, v_{i}, h\right\}$ and the action functional
\begin{eqnarray}
\label{action}
    S(\Phi) &=& 
    v_{i}^{\prime}\,D_F \,v_{i}^{\prime} + v_{i}^{\prime}\left\{ -\nabla_t  
    +\nu_{0} \partial^{2}  \right\} v_{i}
    \\ \nonumber
    &+&
    \frac{1}{2}h^{\prime}h^{\prime} + h^{\prime}\left\{-\nabla_t h + \kappa_0 \partial^2 h + \frac{1}{2}\lambda_0(\partial h)^2 \right\}. 
\end{eqnarray}
Here $D_F$ is the correlation function~(\ref{D}) and all the needed  summations over repeated indices and integrations over $\{ t,  {\bf x}\}$ are implied, 
for example,
\begin{equation}
\label{quadlocal2}
{v_i^{\prime}}\,\partial_t{v_i}= \sum_{i=1}^d\,
\int dt\int d^d{\bf x} \, v_i^{\prime}(t,  {\bf x})\,\partial_t v_i(t,  {\bf x}). 
\end{equation}

Analysis of UV divergences 
shows that the model~(\ref{action}) is multiplicatively renormalizable with the following renormalized action (in the symbolic notation):
\begin{eqnarray}
 S_R(\Phi) = v_{i}^{\prime} \left\{ g_1\mu^y \nu^3 k^{2+\varepsilon-y}+ Z_1 g_2 \,\mu^{\varepsilon}\nu^3 k^{2} \right\} v_{i}^{\prime}+v_{i}^{\prime}\left\{-\nabla_{t} + Z_2 \nu \partial^{2} \right\} v_{i}
   \nonumber \\+\frac{1}{2} Z_3 h^{\prime} h^{\prime}
+h^{\prime}\left\{-\nabla_{t} h  + Z_4\, w \nu \partial^{2} h+\frac{1}{2} Z_5 g_3\, \mu^{\varepsilon/2}\nu^{3/2}(\partial h)^{2}\right\}.
\end{eqnarray}
Here the bare parameters are replaced with their renormalized analogs (without the subscript ``0''), while the momentum reference scale $\mu$ is an additional parameter of the renormalized theory. 
The first term is not renormalized being non-local, while the terms with $\nabla_t$
are not renormalized owing to the Galilean symmetry.

The renormalization constants $Z_1$~-- $Z_5$ are chosen to absorb the UV divergences;
they are related to the renormalization constants of the fields and parameters as follows:
\begin{eqnarray}
\label{ZZ}
Z_{g_1}= Z_{2}^{-3}, \quad
Z_{g_2}=Z_{1} Z_{2}^{-3}, \quad Z_{g_3}=Z_{5} Z_{3}^{1 / 2} Z_{2}^{-3 / 2}, \quad Z_{w}=Z_{4} Z_{2}^{-1}, \\
 Z_{h^{\prime}}=Z_{h}^{-1}=Z_{3}^{1 / 2}, \quad Z_{\nu}=Z_{2}, \quad Z_{v}=Z_{v^{\prime}}=1, \quad {\rm where} \nonumber
 \\
g_{10} = g_1\, \mu^y\, Z_{g_1}, \quad g_{20} = g_2\, \mu^{\varepsilon}\, Z_{g_2}, \quad g_{30} = g_3\, \mu^{\varepsilon/2}\, Z_{g_3}, \quad w_0 = w\, Z_{w}.
\label{ZZ1}
\end{eqnarray}
For brevity, we omit explicit one-loop expressions for the renormalization constants and the corresponding calculation; detailed presentation of similar calculations can be found, e.g., in~\cite{AKL}.

\section{RG analysis and the fixed points of the RG equations \label{rgan}}

The RG analysis allows one to establish possible types of IR asymptotic behaviour of the correlation (Green) functions; see, e.g.,~\cite{Vasiliev} for detailed discussion.
The key role is played by the differential operator $ \widetilde{\cal D}_{\mu} = \mu\partial_{\mu}$ at fixed bare parameters. For the model~(\ref{action}), it is expressed in the renormalized variables as follows:
\begin{equation}
\widetilde{\cal D}_{\mu}
= {\cal D}_{\mu} + \beta_{g_1}\partial_{g_1} +
\beta_{g_2}\partial_{g_2} +
\beta_{g_3}\partial_{g_3}+
\beta_{w}\partial_{w} -
\gamma_{\nu}{\cal D}_{\nu},
\label{RG2}
\end{equation}
where we have written ${\cal D}_{s} \equiv s\partial_{s}$ for any variable
$s$. 

The anomalous dimension $\gamma_{e}$ of a certain parameter $e$
is defined as
$\gamma_{e}= \widetilde{\cal D}_{\mu} \ln Z_e$. 
The $\beta$ functions for all the coupling constants $g_i =\{g_1, g_2, g_3, w\}$
are 
defined as $\beta_{g} = \widetilde {\cal D}_{\mu}g$  and read [see~(\ref{ZZ})]
\begin{eqnarray}
\beta_{g_1} &= g_1(-y-\gamma_{g_1}),& \quad
\beta_{g_2} = g_2 (-\varepsilon-\gamma_{g_2}),
\nonumber \\
\beta_{g_3} &= g_3\left(-\varepsilon/2-\gamma_{g_3}\right),&
\quad 
\beta_{w} = -w \gamma_w.
\label{betagw}
\end{eqnarray}
The one-loop expressions for the anomalous dimensions have the forms:
\begin{eqnarray}
\gamma_1 = \frac{({ g_1}+{ g_2})^2}{32\pi{ g_2}}, \quad
\gamma_2 = \frac{({g_1}+{g_2})}{32\pi}, \quad
\gamma_3 = \frac{{g_3}^2}{16\pi w^3},  \nonumber
\\ 
\gamma_4 = \frac{({ g_1}+{ g_2})}{8\pi w(w+1)}, \quad
\gamma_5 = \frac{({ g_1}+{ g_2})}{8\pi w(w+1)}.
\end{eqnarray}
It follows from~(\ref{ZZ}) that the anomalous dimensions of the coupling constants, the fields and the parameters are:
\begin{eqnarray}
\gamma_{g_1} = -3 \gamma_2; \quad
\gamma_{g_2} = \gamma_1-3\gamma_2; \quad
\gamma_{g_3} = \gamma_5-\frac{3}{2}\gamma_2+\frac{1}{2}\gamma_3;\nonumber
\\
\gamma_w = \gamma_4-\gamma_2; \quad
\gamma_{h}= -\gamma_{h^{\prime}}= -\gamma_{3}/2, \quad \gamma_{v} = \gamma_{v{\prime}}=0, \quad
\gamma_{\nu} = \gamma_{2}.
\label{gadf}
\end{eqnarray}
The one-loop expressions for the $\beta$ functions are as follows:\footnote{These expressions, and hence all the subsequent results, are in agreement with those derived earlier for various special cases in~\cite{FNS,AKL,HoNa,AHKV}.}
\begin{equation}
\beta_{g_1} = g_1  \left\{-y + \frac{3\left({g_1}+{g_2}\right)}{32\pi}\right\},
\label{bg1}
\end{equation}
\begin{equation}
\beta_{g_2}= g_2  \left\{-\varepsilon - \frac{({g_1}+{g_2})^2}{32\pi{ g_2}} + \frac{3({ g_1} + { g_2})}{32\pi} \right\},
\label{bg2}
\end{equation}
\begin{equation}
\beta_{g_3}= g_3  \left\{-\frac{\varepsilon}{2}
-\frac{({g_1}+{ g_2})}{8\pi w(w+1)} + 
\frac{3({ g_1}+{ g_2})}{64\pi} - 
\frac{{ g_3}^2}{32\pi w^3}\right\},
\label{bl}
\end{equation}
\begin{equation}
\beta_w =\frac{-w  ({ g_1}+{ g_2})}{2\pi}  \left\{\frac{1}{4 w(w+1)} -  \frac{1}{16}\right\}.
\label{bw}
\end{equation}

Possible asymptotic scaling regimes of the model are determined by the fixed points of the RG equations.
The coordinates of the fixed points are determined by the zeroes of the $\beta$ functions. The type of a fixed point is determined by the matrix 
$\Omega_{ij} = \partial_i \beta_j$, where $g_i =\{g_1, g_2, g_3, w\}$ is the full set of couplings and $\beta_i$ is the full set of $\beta$ functions. For an IR attractive point the matrix $\Omega$ is non-negative, i.e., all its eigenvalues $\lambda_i$ have non-negative real parts.

Since the scalar field does not affect the velocity, the functions $\beta_{g_1}$
and $\beta_{g_2}$ do not depend on $g_3$ and $w$ and can be studied separately. They have three fixed points, corresponding to simple diffusion of the velocity field (regime number 1), fluid in thermal equilibrium (regime number 2) and turbulent fluid (regime number 3). In the full model, each of them splits into two points: one with $g_3=0$ (the KPZ nonlinearity is irrelevant) and another with $g_3\ne0$; we label them by A and B, respectively. Thus, we have six fixed points.

The first pair of fixed points has the coordinates $g_{1*} = g_{2*}=0$ and corresponding eigenvalues $\lambda_1 = -y$, $\lambda_2 = -\varepsilon$. It includes the points 1A and 1B:

1A. The fixed point with the coordinate $g_{3*}=0$ and arbitrary $w_{*}$ (so it is rather a line of fixed points); two remaining eigenvalues are $\lambda_3=-{\varepsilon}/{2}$ and $\lambda_4 = 0$. The eigenvalues $\lambda_1, \lambda_2$ and $\lambda_3$ are positive when $y<0$ and $\varepsilon<0$. This is a regime of ordinary diffusion both for the velocity and scalar fields.

1B. The fixed point with the coordinate ${g}_{3*}^2=-16\pi \varepsilon w_{*}^3$ and arbitrary $w_{*}$; two remaining eigenvalues are $\lambda_{3}=\varepsilon$ and $\lambda_{4} = 0$. Due to the equality $\lambda_{2} = -\lambda_{3}$ this fixed point is IR attractive only on the half-line $\varepsilon=0, y<0$. Here ${g}_{3*}=0$ which again leads to a simple diffusion. 

The second pair of the fixed points has the coordinates $g_{1*} = 0$, $g_{2*}=16\pi\varepsilon$ and corresponding eigenvalues $\lambda_{1} = -y+3\varepsilon/2$, $\lambda_{2} = \varepsilon$. This regime corresponds to a fluid in thermal equilibrium. The pair includes points 2A and 2B:

2A. The fixed point with the coordinates $g_{3*}=0$ and $w_{*}=(\sqrt{17}-1)/2$; two remaining eigenvalues are $\lambda_{3}=-\varepsilon/4$ and  $\lambda_{4} = \varepsilon/2+8\varepsilon(1+\sqrt{17})^{-2}$. This fixed point is also IR attractive only on the half-line $\varepsilon=0, y<0$. 

2B. The fixed point with the coordinates  ${g}_{3*}^2=- (\sqrt{17}-1)^3\pi\varepsilon$ and $w_{*}=(\sqrt{17}-1)/2$; two remaining eigenvalues are $\lambda_{3}=\varepsilon/2$ and $\lambda_{4} = \varepsilon/2+8\varepsilon/(1+\sqrt{17})^{-2}$. When $\varepsilon>0$ and $y<3\varepsilon/2$  all the eigenvalues are positive and the fixed point is IR attractive. This is a regime where both the advection and the KPZ nonlinearity are relevant.  However, ${g}_{3*}^2$ is negative, which is a typical feature of the perturbative RG approach to the KPZ model that requires a careful physical interpretation.

3. The third pair of fixed points points 3A and 3B has the coordinates
\begin{eqnarray}
g_{1*} = \frac{32\pi}{9} \frac{y (3\varepsilon - 2y)}{\varepsilon - y}, \quad
g_{2*} = \frac{32\pi}{9} \frac{y^2}{y-\varepsilon }. \nonumber
\end{eqnarray} 
The corresponding eigenvalues are
$\lambda_{1,2}= -\varepsilon/2 + 2y/3 \pm \sqrt{9 \varepsilon^2 + 12\varepsilon y - 8y^2}/6$. This is a regime with turbulent motion of the environment (the non-local term in~(\ref{D}) is relevant as $ g_{1*}\ne 0$).

3A. The fixed point with the coordinates $g_{3*}=0$ and $w_*= (\sqrt{17}-1)/2$; two remaining eigenvalues are $\lambda_3= -\varepsilon/2 + y/6$ and $\lambda_4= (17- \sqrt{17}) \,y/24$. 
This is a regime of pure turbulent scalar field advection. This point is IR stable for  $3\varepsilon/2 < y$. 
One can check that for  $\lambda_3 > 0$ the expression $(9 \varepsilon^2 + 12\varepsilon y - 8y^2)$ is negative, so that the eigenvalues  
$\lambda_{1}$ and $\lambda_{2}$ are complex. Therefore, this point is a focus in the $g_1$~--~$g_2$ plane.

3B. The fixed point with the coordinates 
$g_{3*}^2=2\pi \left(y/3-\varepsilon\right)(\sqrt{17}-1)^3$ and
$w_*= (\sqrt{17}-1)/2$; two remaining eigenvalues are 
$\lambda_3= \varepsilon - y/3$ and $\lambda_4= (17- \sqrt{17}) \,y/24$. For $3\varepsilon/2 < y < 3\varepsilon$ the fixed point is IR attractive.  This a regime where both the turbulent advection and the KPZ nonlinearity are relevant. However, $g_{3*}^2$ is negative for these values of $\varepsilon$ and $y$.

The general stability pattern of the fixed points in the plane $(y,\varepsilon)$ is shown in Fig.~1.
The existence of IR attractive solutions of the RG equations leads
to the existence of the scaling behavior of correlation functions. 
The critical exponents in the scaling relation~(\ref{scaling}) corresponding to these scaling regimes can be calculated in a standard fashion; see, e.g.,~\cite{Red,Vasiliev} for general scheme and~\cite{AKL,Us} for similar models. For the most interesting from physical point of view point 3A, corresponding to situation when $y\to4$  (large-scale stirring) and $d=2$ or $3$, they read
\begin{eqnarray}
z=2-y/3, \quad \chi=\varepsilon/2-y/6.
\label{kuku}
\end{eqnarray} 
These expressions are exact, that is, they have no higher-order corrections in $y$ and 
$\varepsilon$. The absence of corrections follows from direct calculations together with the fact that $g_{3*}=0$ for this point.

\begin{figure}[h!]
\center{\includegraphics[width=7cm]{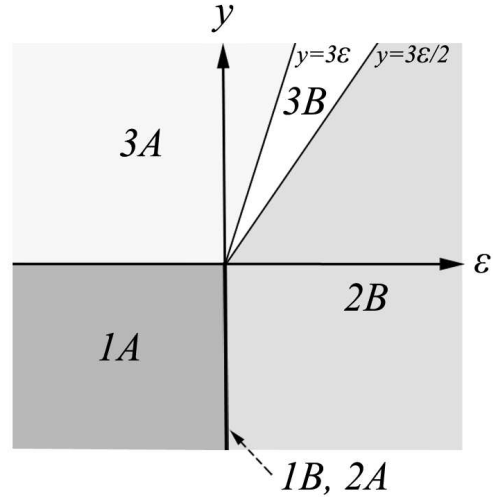}
\caption{Regions of IR attraction of the fixed points in the model~(\ref{action}).}}
\label{fig}
\end{figure}

\section{Conclusion \label{conc}}

We studied effects of turbulent environment on the scaling behaviour of a randomly growing surface. The latter was described by the KPZ model~(\ref{KPZ}) which made our findings applicable to a wide class of non-equilibrium critical systems. The advecting field was modelled by the stochastic NS equation~(\ref{NS}) with a power-like correlation function of the stirring force consisting of two terms~(\ref{D}): one term is non-local and, for $y\to4$, represents the input of energy by  largest-scale motions, while the second term is local and was required for renormalizability near $d=2$. 
As a byproduct, such choice of the correlation function allows one to consider both the case of a turbulent motion and the case of a fluid in thermal equilibrium. 

The field theoretic RG analysis was applied in the leading order of perturbation theory (one-loop approximation). It was obtained that the model reveals six possible regimes of IR scaling behaviour associated with the six fixed points of the RG equations. 

The diagram 
of the fixed points stability regions (see Fig.~1) shows that there are neither gaps  nor overlaps between different regions. However, this can be an artefact of the one-loop approximation and the gaps or overlaps can appear in higher-order approximations \cite{AHKV}.

It was found that for the case of a turbulent fluid ($y\to4$) 
and the most interesting values of the spatial dimension ($d=2$ or $3$, i.e.,  $\varepsilon=0$ or $-1$) the effects of the KPZ nonlinearity are ``washed away'' by the flow and the IR behaviour is described by the regime of pure turbulent advection (point 3A) with exactly known scaling exponents~(\ref{kuku}) in the scaling representation~(\ref{scaling}). 
It is also worth mentioning that similar results were derived earlier for a simpler velocity statistics (the Kazantsev--Kraichnan ensemble), see~\cite{Us}. In this sense the obtained result seems to be a feature of KPZ equation itself rather than an artefact of the model under consideration.

One can hope that these one-loop results will not change qualitatively when the higher-order corrections are taken into account. This statement is supported by the two-loop calculation~\cite{AHKV} for the NS equation with the stirring force~(\ref{D}) employed in our paper. However, the two-loop analysis of the full model is welcome and it is an interesting problem for the future. 

Another important question is the fate of the strong-coupling, essentially nonperturbative fixed point of the KPZ model~(\ref{KPZ1})~-- (\ref{covar}) whose existence was hypothesized in the phenomenology and strongly supported by the functional RG~\cite{Canet}~--~\cite{Canet4}.
If that point indeed exists, it should be necessarily present in our model~(\ref{covar})~-- (\ref{D}). At the same time, it can become unstable with respect to the turbulent advection (i.e., it may be the saddle type point; as is shown to happen with the perturbative KPZ point). This means, that the resulting IR behaviour of the system is really governed either by pure turbulent advection (regime 3A) founded by our analysis or by a new  strong-coupling fixed point invisible by standart RG technique and corresponding to the regime where both the nonlinearity and advection are simultaneously important.
In order to resolve this dilemma, one has to apply the functional RG to our model, which clearly is a highly difficult task already for the pure NS equation itself~\cite{Canet5}.

It also would be interesting to study the feedback of the KPZ scalar field on the dynamics of the advecting velocity (``active scalar''). The RG study of this problem for the linear advection-diffusion equation shows that the active term in the NS equation appears IR irrelevant for the turbulent case \cite{Nandy} and for the case thermal equilibrium \cite{Maria}. One can hope that inclusion of the KPZ nonlinearity will produce a fully nontrivial scaling behaviour where the velocity dynamics is affected by the scalar field. This work is already in progress.

\begin{ack}

The reported study was funded by RFBR, project number~20-32-70139.
The work by N~V~Antonov and P~I~Kakin was also supported by the Foundation for the Advancement of Theoretical Physics and Mathematics ``BASIS.''

The authors thank the Organizers of the 8$^{\rm th}$ International Conference on New Frontiers in Physics (Crete, ICNFP, 21~-- 29 August 2019) 
for possibility to present results of the present study.

\end{ack}


\end{document}